\documentclass{PoS}

\title{Twisted reduction in large N QCD with two adjoint Wilson fermions}

\ShortTitle{Twisted reduction in large N QCD with two adjoint Wilson fermions}

\author{Antonio Gonz\'alez-Arroyo$^{ab}$ and \speaker{Masanori Okawa}$^c$\\
\llap{$^a$}Instituto de F\'{\i}sica Te\'orica UAM/CSIC\\
\llap{$^b$}Departamento de F\'{\i}sica Te\'orica, C-15 \\
       Universidad Aut\'onoma de Madrid, E-28049--Madrid, Spain \\
\llap{$^c$}Graduate School of Science, Hiroshima University\\
Higashi-Hiroshima, Hiroshima 739-8526, Japan\\
E-mail: \email{antonio.gonzalez-arroyo@uam.es}, \email{okawa@sci.hiroshima-u.ac.jp}}

\abstract{The twisted reduced model of large $N$ QCD with two adjoint Wilson fermions is 
studied numerically using the Hybrid Monte Carlo method. This is the one-site model, 
whose large $N$ limit (large volume limit) is expected to be conformal or nearly conformal. 
The symmetric twist boundary condition with flux $k$ is used. $k$=0 corresponds to periodic 
boundary conditions. It is shown that the quark mass and $N$ dependencies of the model 
with non-vanishing $k$ differ significantly from those of the $k$=0 model. A preliminary result
for the string tension calculated at $N$=289 is presented.  The string tension seems to vanish 
as the physical quark mass decreases to zero in a way consistent with the theory being 
governed by an infrared fixed point with  $\gamma_* = 0.8 \sim 1.2$.}

\FullConference{The 30th International Symposium on Lattice Field Theory\\
		 June 24 - 29,  2012\\
		 Cairns, Australia}

\begin{document}

\section{Introduction}

\vspace{- 0.2 cm}

Recently the SU(N) lattice gauge theories with two adjoint Wilson
fermions have received much attention, since their massless limits  are expected to be 
conformal or nearly conformal\cite{DD}.   In fact the first two coefficients 
of the perturbative $\beta$ function expressed in terms of the 't Hooft coupling are given by
$b_0=(11-4N_f)/24\pi^2$ and $b_1=(17-16N_f)/192\pi^4$, with $N_f$ the
number of adjoint quark flavors. This  would  imply the existence of 
an infrared fixed point if $17/16 < N_f < 11/4$ irrespective of the value of gauge group $N$. 
In particular, the SU(2) theory has been studied by many authors, and now there is a consensus that 
this theory is conformal and the chiral condensate anomalous dimension at the
infrared fixed point is approximately $\gamma_* \sim 0.3$\cite{DD}.

The purpose of the present talk is to study the large $N$ lattice gauge theory with two adjoint 
Wilson fermions directly by means of the twisted space-time reduced
model\cite{TEK,TEK2}.  This is a one-site model given by four SU(N)
matrices only. There is a close connection between this model with
$N=L^2$, with the usual SU(N) theory on a $L^4$ space-time lattice.
In the large $L$ limit this model should reproduce the corresponding large $N$
infinite volume theory.  It should be noted that the use of the
space-time reduced model has a great advantage in the treatment of the dynamical quark effects. 
It is almost impossible to make a simulation with both a large SU(N)
internal group and large volume $V$, since the number of the degrees of freedom is 
too large.  On the other hand,  the number of degrees of freedom is
significantly reduced, by a factor $V$, in the space-time reduced model, and the study of the 
dynamical quark effects is manageable in this theory.

It is widely advocated~\cite{KUY,AHUY,BKS} that the introduction of the dynamical adjoint Wilson fermions is enough to 
restore the Z$^4$(N) symmetry which is needed for the reduction idea
to hold\cite{EK}, and is broken\cite{QEK} in the original Eguchi-Kawai model 
(EK model)\cite{EK}.  Our results also confirm that, with the inclusion
of the dynamical adjoint fermions, the Z$^4$(N) symmetry is indeed unbroken for a wide
range of values of $N$ and bare quark masses.  At the same time, however, we also find that this theory 
has a very strong $N$ dependence, making it practically useless for
extracting the large N behavior. Our proposal is to apply  twisted
boundary conditions in the formulation of the space-time reduced
model.  For the pure gauge theory, we have recently demonstrated that the twisted 
reduction (the twisted Eguchi Kawai model) works quite well, allowing
a determination of the continuum string tension\cite{TEK3,TEK4}.  We will show that the model with
dynamical adjoint Wilson fermions and twisted boundary conditions has
significantly less $N$ dependence than its untwisted counterpart.  
This has allowed us to present a preliminary calculation of the large
N string tension using $N$=289. We show that the string tension seems to vanish as we decrease
the quark mass towards zero, as predicted by the existence of an infrared
fixed point. The rate of decrease points towards a large anomalous
dimension  $\gamma_* = 0.8 \sim 1.2$ at this point.     

\vspace{- 0.3 cm}
   
\section{Formulation}

\vspace{- 0.2 cm}

We consider SU(N) gauge group with $N=L^2$, $L$ being a positive integer.  Then the action of the model is given by

\vspace{- 0.3 cm}

\begin{equation}
 S= b N \sum_{\mu \ne \nu =1}^4 {\rm Tr} \left(Z_{\mu\nu}U_\mu U_\nu U_\mu^\dagger U_\nu^\dagger \right) 
    +\sum_{j =1}^{N_f} {\bar \Psi}_j D_W \Psi_j . 
\end{equation}

\noindent
$b$ is the inverse 't Hooft coupling $b=1/g^2N$.  $U_\mu$ are four
SU(N) matrices and $\Psi_j$ are the fermion fields  in the adjoint
representation of color SU(N) for $N_f=2$.  The twist tensor

\vspace{- 0.2 cm}

\begin{equation}
 Z_{\mu\nu} = \exp \left( k {2\pi i \over L} \right), \ \ \ Z_{\nu\mu}=Z_{\mu\nu}^*, \ \ \ \mu>\nu 
\end{equation}

\noindent
introduces  symmetric twisted boundary conditions to the space-time reduced model specified by the flux $k$.
The weak coupling analysis demands that $k$ and $L$ are co-prime for
the model to work well in that limit\cite{TEK2}. 
The $k$=0 case corresponds to periodic boundary conditions.  The Wilson Dirac matrix is given by

\vspace{- 0.3 cm}

\begin{equation}
 D_W = 1- \kappa \sum_{\mu=1}^4 \left[ (1-\gamma_\mu)U_\mu^{adj} +  (1+\gamma_\mu)U_\mu^{adj \dagger} \right].
\end{equation}

\noindent
$U_\mu^{adj}$ are the link variables acting on the adjoint fermions in the color $(N,{\bar N})$ representation 
as $U_\mu^{adj}\Psi_j = U_\mu \Psi_j U_\mu^{\dagger}$.

We have studied the previous model by means of the Hybrid Monte Carlo
method.  The simulation has been done on Hitachi SR16000 super computer with one node
having 32-cores IBM power7 and peak speed of 980 GFlops/sec.  Thanks
to the Hitachi system engineers, our codes are highly optimized for
SR16000 and the sustained seed is 300-600 GFlops/sec depending on the
values of $N$  and $\kappa$.

To study the $N$ dependence of the model precisely, we have performed
simulations at  $N$= 25, 49, 81,
121, 169, 225, 289, which correspond to all odd integer values  of $L$ from 5
to 17. It is known that part of the finite $N$ effects for our twisted model 
correspond to the finite volume effects in  the usual lattice  theory
on an $L^4$  lattice\cite{TEK}: Hence, we based our results on Wilson loops 
$W(R,T)$ up to $R$=$T$=$(L-1)/2$.  For $N$=289 this corresponds to $R$=$T$=8.

\vspace{- 0.3 cm}
   
\section{$\kappa$ and $N$ dependence}

\vspace{- 0.2 cm}
   
\begin{figure}[b]
\centering
\includegraphics[width=.6\textwidth]{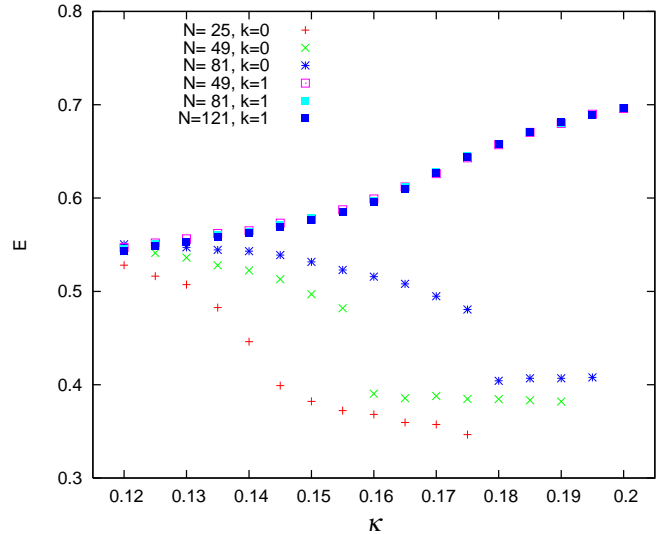}
\caption{$\kappa$ dependence of $E$ for various $N$ at $b$=0.35 and $k$=0 and 1.}
\label{fig1}
\end{figure}

In fig. 1 we show the $\kappa$ dependence of the average of plaquette 
$E=<Z_{\mu\nu}U_\mu U_\nu U_\mu^\dagger U_\nu^\dagger>$ at $b$=0.35.
For periodic boundary conditions $k$=0, we show the results for $N$=25,49 and 
81.  As $\kappa$ increases $E$ decreases, which is quite unnatural since as $\kappa$ increases one usually expects
to have larger dynamical quark effects which tends to order the system
and, thus, to increase the internal energy $E$.  Furthermore, there exists an abrupt change in
$E$ at intermediate values of $\kappa$. However, the position of the
abrupt change depends   significantly on the value of $N$.  This
change is not related to Z$^4$(N) symmetry breaking, which was found to 
remain unbroken for the set of simulation parameters presented in this talk.
The large $N$ dependence suggests that it is related to some finite
$N$ artifact. The results for twisted boundary conditions and $k$=1
are shown for $N$=49, 81 and 121.  In this case $E$ increases as we 
increase $\kappa$ as expected, and the $N$ dependence looks rather small 
within the rough scale of the plot.  

\begin{figure}
\centering
\includegraphics[width=.6\textwidth]{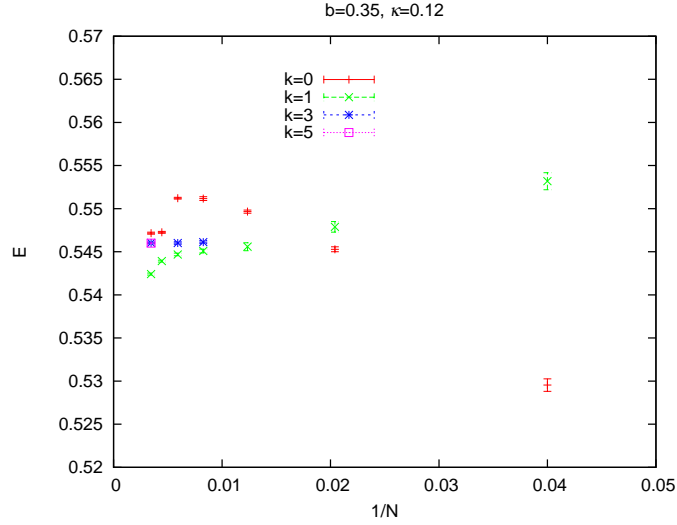}
\caption{$N$ dependence of $E$ for various $k$ at $b$=0.35 and $\kappa$=0.12.}
\label{fig2}
\end{figure}

A more precise study of the $N$ dependence is given in fig. 2, where
$E$ is plotted as a function of $1/N$ for  $b$=0.35, $\kappa$=0.12
and various values of $k$.  As we have stated previously, at $\kappa$=0.12 there 
is no Z$^4$(N) symmetry breaking.  However, the results at finite value of $\kappa$ seem to be strongly affected 
by the nearby presence of the Z$^4$(N) symmetry breaking point at $\kappa$=0, namely the reduced model without
fermions.   We know that for the pure gauge theory  ($\kappa$=0) the
breaking occurs as follows\cite{QEK,IO,TV,AHHI,TEK2}:

 
\vspace{ 0.1 cm}

$\bullet$ $k$=0, Z$^4$(N) symmetry is broken for all $N>0$. 

$\bullet$ $k$=1, Z$^4$(N) symmetry is broken for $N>100$. 

$\bullet$ $k$=3, Z$^4$(N) symmetry is broken for $N>784$.

$\bullet$ $k$=5, Z$^4$(N) symmetry is expected to be broken for $N>2200$.
\vspace{ 0.1 cm}
 
Now let us look at fig. 2 more carefully.  The data with $k$=0 have,
again, a large $N$ dependence. For $N \le $121, this dependence seems
linear in  $1/N$.  However, such a linear extrapolation to the large $N$
value of $E$ is quite misleading, since the behavior of $E$ suddenly changes for $N$>121.
The data with $k$=1 have less $N$ dependence.  However, as we increase
beyond $N$=121, $E$ also starts to decrease. We think this behavior is closely related to 
the Z$^4$(N) symmetry breaking at $\kappa$=0.  If this interpretation is correct, we should expect 
a similar pattern for $k$=3(5), with a smooth $N$ dependence up to
$N \le 784$($N \le 2200$).  In fact, we observe that  the two data points at $N$=289 for $k$=3 and 5 are quite consistent. 
In fig. 3 and 4, we display the same type of information but for
$(b,\kappa)$ values of (0.36, 0.12) and (0.35, 0.14) respectively. 
Details are different, but the overall trends are the same as in Fig. 2.

\begin{figure}[htbp]
 \begin{minipage}{0.5\hsize}
  \begin{center}
   \includegraphics[width=70mm]{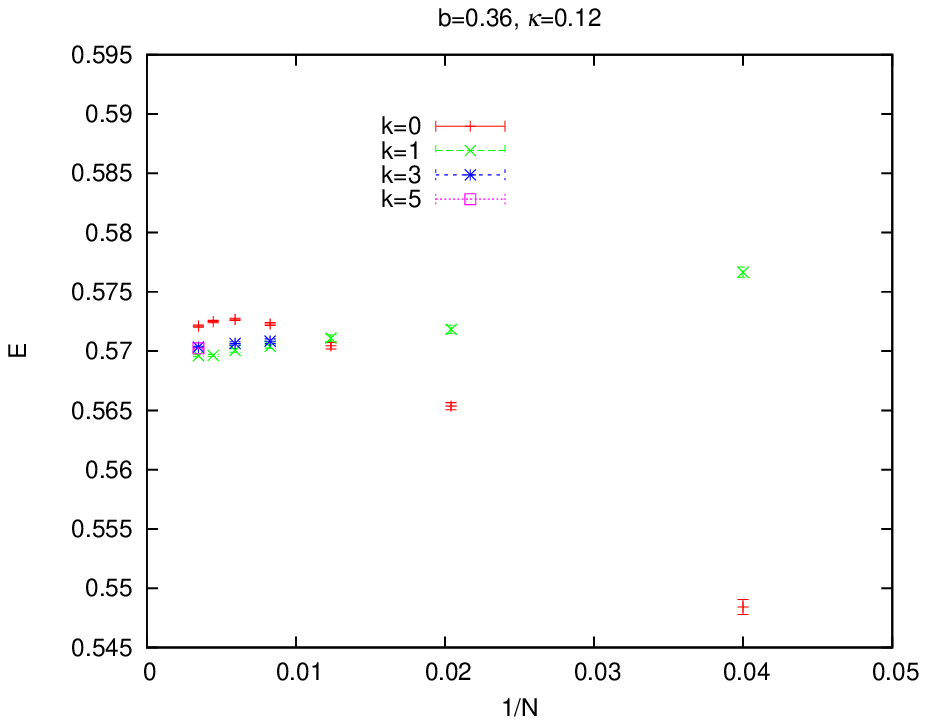}
  \end{center}
  \caption{$N$ dependence of $E$ at $b$=0.36 and $\kappa$=0.12.}
  \label{fig3}
 \end{minipage}
 \begin{minipage}{0.5\hsize}
  \begin{center}
   \includegraphics[width=70mm]{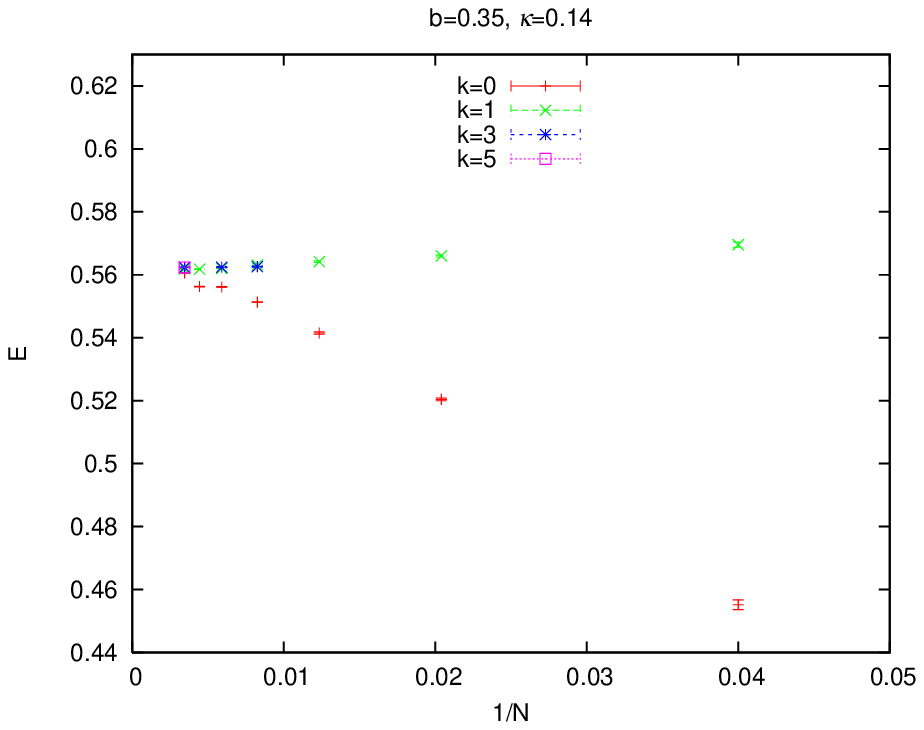}
  \end{center}
  \caption{$N$ dependence of $E$ at $b$=0.35 and $\kappa$=0.14.}
  \label{fig4}
 \end{minipage}
\end{figure}

\vspace{- 0.3 cm}
 
\section{String tension at $N$=289.} 

\vspace{- 0.2 cm}

Based on the results of the previous section, we made an attempt at a
determination of the string tension for $N$=289 and $k$=5 and various
values of $\kappa$. We used the same procedure as used for the pure
gauge reduced model in order to extract the string tension from 
rectangular  $R \approx T$ Wilson loops $W(R,T)$~\cite{TEK3,TEK4}. 
Namely, we extract the string tension from Creutz ratios formed out of 
Wilson loops with smeared links. The  Ape smeared links, defined as

\vspace{- 0.2 cm}

\begin{equation}
 U_\mu^{smeared}={\rm Proj}_N \left[ U_\mu + c \sum_{\nu \ne \mu}(Z_{\nu \mu} U_\nu U_\mu U_\nu^\dagger 
                + Z_{\mu \nu} U_\nu^\dagger U_\mu U_\nu) \right], 
\end{equation}    

\noindent
are used to enhance the signal to noise ratio.
 ${\rm Proj}_N$ stands for the operator for projection onto the SU(N) matrices. 

\begin{figure}[htbp]
 \begin{minipage}{0.5\hsize}
  \begin{center}
   \includegraphics[width=70mm]{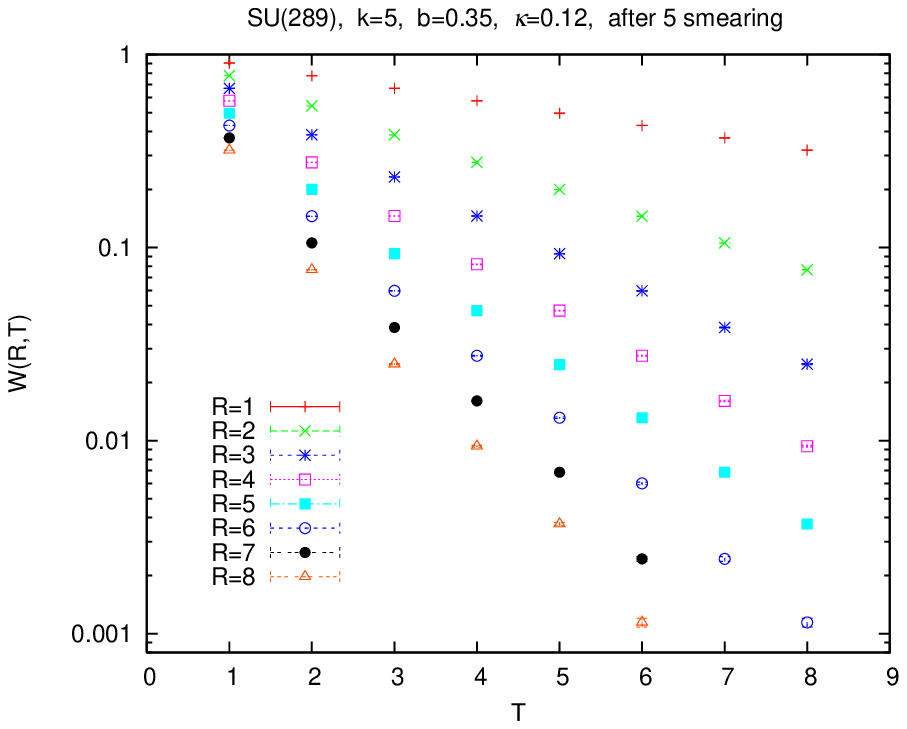}
  \end{center}
  \caption{Wilson loop after 5 smearing.}
  \label{fig5}
 \end{minipage}
 \begin{minipage}{0.5\hsize}
  \begin{center}
   \includegraphics[width=70mm]{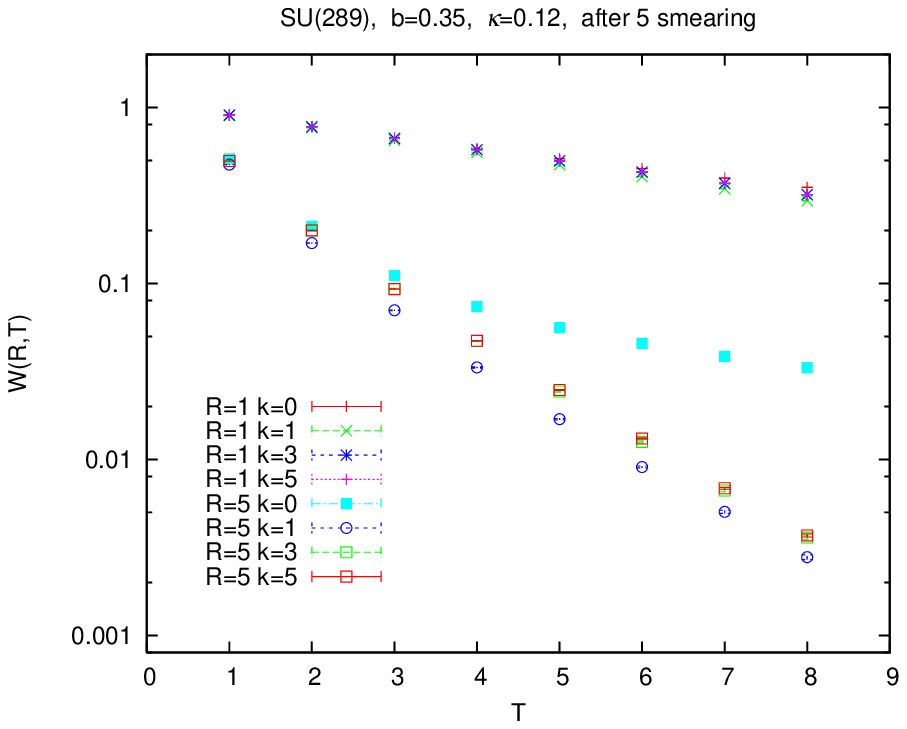}
  \end{center}
  \caption{Wilson loop for various values of $k$.}
  \label{fig6}
 \end{minipage}
\end{figure}

In fig. 5, we show Wilson loops $W(R,T)$ after 5 smearing with  smearing parameter $c$=0.1.
We observe a clear signal up to loop size 8, which according to our
previous comment, corresponds to slighly less than half the effective
linear lattice size $L$=17 for our value of  $N$=289=$17^2$.

In fig. 6, we compare Wilson loops calculated with different $k$
values.  In this rough scale, Wilson loops with $R$=1, $W(R=1,T)$, look consistent 
for all values of $k$=0,1,3 and 5.  On the other hand, the situation is quite 
different for $R$=5.  The results with $k$=3 and 5 coincide with each
other, giving what we think are the correct values.
It is amazing that the results with $k$=0 are so different from those of $k$=3 and 5 as $T$ increases.  
The Wilson loops $W(R=5,T)$ with $k$=1 also have smaller values than
those with $k$=3 and 5. We believe that this is also related to the
instabilities occurring for $k=1$ at $N>100$ with $\kappa$=0. 

\begin{figure}[htbp]
 \begin{minipage}{0.5\hsize}
  \begin{center}
   \includegraphics[width=70mm]{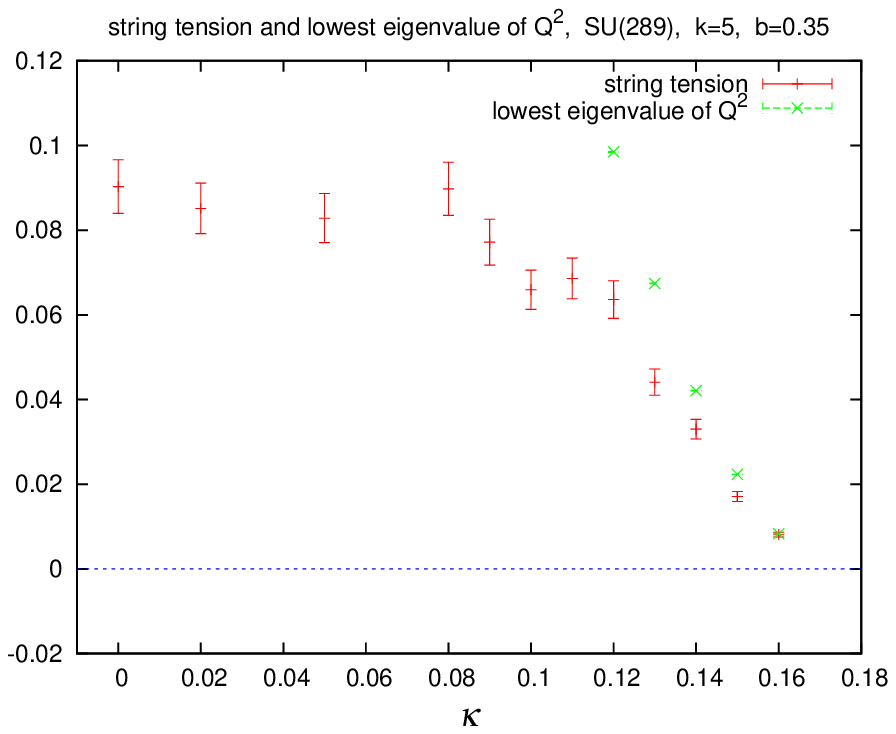}
  \end{center}
  \caption{$\sigma$ and lowest eigenvalue of $Q^2$ at $b=0.35$.}
  \label{fig7}
 \end{minipage}
 \begin{minipage}{0.5\hsize}
  \begin{center}
   \includegraphics[width=70mm]{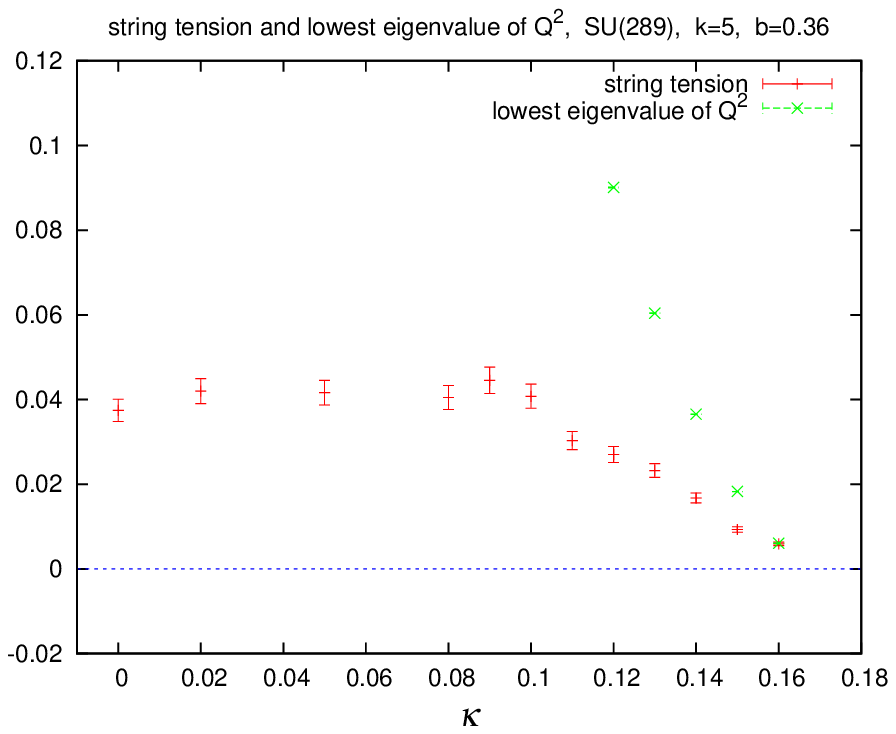}
  \end{center}
  \caption{$\sigma$ and lowest eigenvalue of $Q^2$ at $b=0.36$.}
  \label{fig8}
 \end{minipage}
\end{figure}

In fig. 7, we show with red symbols our preliminary results for the extracted string tension $\sigma$ as a function 
of $\kappa$ calculated at $N$=289, $k$=5, $b$=0.35.  The value at $\kappa$=0 is obtained from the Twisted Eguchi Kawai model without fermions.
As we increase $\kappa$, the string tension decreases and seems to vanish around $\kappa$=0.17. 

Although at present we have not calculated any fermionic spectrum, it is straightforward to calculate the lowest 
eigenvalue of the positive square hermitian Wilson-Dirac 
operator $Q^2=(\gamma_5 D_W)^2$, which should be related to quark mass square. In fig. 7, we also plot
the lowest eigenvalue of $Q^2$ with green symbols.  
It seems that the string tension and the lowest eigenvalue of $Q^2$ vanish at the same point.  

If the massless theory is governed by the infrared fixed point at
which the theory is conformal, all physical quantities having dimension of mass square
should behave as

\vspace{- 0.2 cm}

\begin{equation}
 \sigma \sim \left( {1 \over \kappa} - {1 \over \kappa_c} \right)^{2 \over 1+ \gamma_*},
\end{equation}    
      
\noindent
with $\gamma_*$ the chiral condensate anomalous dimension at the infrared fixed
point.  We assume that the quark mass is proportional to $(1/\kappa - 1/\kappa_c)$.   Fitting the $b=0.35$ data in the range $\kappa$=0.12 - 0.16 with a common vanishing point
$\kappa=\kappa_c$ for the string tension and  the lowest eigenvalue of
$Q^2$,  we get a rather large  estimate for  $\gamma_* = 0.81(8)$.
This is considerably larger than  the value  obtained by other authors
for $N$=2\cite{DD}.  It should be noted,  however, that our data of the string tension are quite preliminary. 
In particular, it is based on  small number of  configurations of
order 100 with a rather low global acceptance ratio in 
HMC at $\kappa$=0.15 and 0.16.  

In fig. 8, we repeat the analysis for the data at $b=0.36$.  Again, requiring that the string tension and the 
lowest eigenvalue of $Q^2$ vanish at the same point, our data for the string tension
in the range  $\kappa$=0.12 - 0.15 give  $\gamma_* = 1.17(21)$.

\vspace{- 0.3 cm}

\section{Conclusions}

\vspace{- 0.2 cm}

We have demonstrated that the twisted reduced model of large $N$ QCD with two adjoint Wilson fermions works quite well.
The $N$ dependence of the model with twisted boundary conditions
($k\ne$0) is significantly smaller than that of the model with
periodic boundary condition ($k$=0).  The string tension  calculated at $N$=289, $k$=5, 
clearly decreases as we increase $\kappa$ and seems to vanish around $\kappa$=0.17
in a way  consistent with the expectations for an infrared fixed point
with a rather large value for  $\gamma_* = 0.8 \sim 1.2$.

We have started new simulations for small quark mass with reasonable
global acceptance ratio.  We have also launched the simulation of the 
single flavor adjoint Wilson fermion using the Rational Hybrid Monte Carlo method.  
We hope to present new results at the coming lattice conference.

\vspace{  0.2 cm}

A.G-A is supported from Spanish grants FPA2009-08785, FPA2009-09017, CSD2007-00042, HEPHACOS S2009/ESP-1473,
PITN-GA-2009-238353 (ITN STRONGnet) and CPAN CSD2007-00042. M.O is supported in part by Grants-in-Aid for Scientific Research from the Ministry of Education, Culture, Sports, Science and Technology (No 23540310).

The calculation has been done on Hitachi SR16000-M1 computer at High Energy Accelerator
Research Organization (KEK) supported by the Large Scale Simulation Program No.12-01
(FY2011-12). 
 
\vspace{- 0.3 cm}

\end{document}